\documentclass[fleqn,usenatbib]{mnras}

\usepackage{newtxtext,newtxmath}
\usepackage{booktabs}
\usepackage{multirow}

\usepackage[T1]{fontenc}

\DeclareRobustCommand{\VAN}[3]{#2}
\let\VANthebibliography\thebibliography
\def\thebibliography{\DeclareRobustCommand{\VAN}[3]{##3}\VANthebibliography}

\usepackage{graphicx}	% Including figure files
\usepackage{amsmath}	% Advanced maths commands
\usepackage{ulem}
\usepackage{subcaption}
\newcommand{\Sec}[1]{Section~\ref{#1}}
\newcommand{\App}[1]{Appendix~\ref{#1}}
\newcommand{\Eq}[1]{Eq.~(\ref{#1})}
\newcommand{\Fig}[1]{Fig.~\ref{#1}}
\newcommand{\Tab}[1]{Tab.~\ref{#1}}

\newcommand{\hGpc}{{\ifmmode{,{\rm h^{-1}{\rm Gpc}}}\else{${\rm h^{-1}}$Gpc}\fi}}

\newcommand{\hMpc}{{\ifmmode{,{\rm h^{-1}{\rm Mpc}}}\else{${\rm h^{-1}}$Mpc}\fi}}

\newcommand{\hkpc}{{\ifmmode{,{\rm h^{-1}{\rm kpc}}}\else{${\rm h^{-1}}$kpc}\fi}}

\newcommand{\hpc}{{\ifmmode{,{\rm h^{-1}{\rm pc}}}\else{${\rm h^{-1}}$pc}\fi}}

\newcommand{\hMsun}{{\ifmmode{\,{\rm h^{-1} {M_{\odot}}}}\else{${\rm{h^{-1}M_{\odot}}}$}\fi}}
\newcommand{\Mstar}{{\ifmmode{,M_{*}}\else{$M_{*}$}\fi}}
\newcommand{\Mhalo}{{\ifmmode{\,M_{\rm halo}}\else{$M_{\rm halo}$}\fi}}
\newcommand{\ltsima}{$\; \buildrel < \over \sim \;$}
\newcommand{\gtsima}{$\; \buildrel > \over \sim \;$}
\newcommand{\lsim}{\lower.5ex\hbox{\ltsima}}
\newcommand{\gsim}{\lower.5ex\hbox{\gtsima}}

\newcommand{\monofonic}{\textsc{monofonIC}}
\newcommand{\music}{\textsc{MUSIC}}
\newcommand{\gizmo}{\textsc{GIZMO}}
\newcommand{\gadgetIII}{\textsc{GADGET-3}}
\newcommand{\ahf}{\textsc{AHF}}
\newcommand{\mergertree}{\textsc{MergerTree}}

\usepackage[dvipsnames]{xcolor}

\title{Tidal adaptive softening and artificial fragmentation in cosmological simulations}

\author[Robert A. Mostoghiu Paun et al.]{
Robert A. Mostoghiu Paun,$^{1,2,3}$\thanks{E-mail: rmostoghiupaun@swin.edu.au}
Darren Croton,$^{1,2,3}$
Chris Power,$^{2,3,4}$
Alexander Knebe,$^{4,5,6}$
\newauthor{}
Adam J. Ussing,$^{1,2,3}$
and Alan R. Duffy$^{1,2,3}$
\\
$^{1}$Centre for Astrophysics and Supercomputing, Swinburne University of Technology, PO Box 218, Hawthorn VIC 3122, Australia\\
$^{2}$ARC Centre of Excellence for Dark Matter Particle Physics (CDM)\\
$^{3}$ARC Centre of Excellence for All Sky Astrophysics in 3 Dimensions (ASTRO 3D)\\
$^{4}$International Centre for Radio Astronomy Research (ICRAR), M468, University of Western Australia, 35 Stirling Hwy, Crawley, WA 6009, Australia\\
$^{5}$Departamento de F\'isica Te\'{o}rica, M\'{o}dulo 15, Facultad de Ciencias, Universidad Aut\'{o}noma de Madrid, 28049 Madrid, Spain\\
$^{6}$Centro de Investigaci\'{o}n Avanzada en F\'isica Fundamental (CIAFF), Facultad de Ciencias, Universidad Aut\'{o}noma de Madrid, 28049 Madrid, Spain
}

\date{Accepted XXX. Received YYY; in original form ZZZ}

\pubyear{2024}

\begin{document}
\label{firstpage}
\pagerange{\pageref{firstpage}--\pageref{lastpage}}
\maketitle

\begin{abstract}
Traditional N-body methods introduce localised perturbations in the gravitational forces governing their evolution. These perturbations lead to an artificial fragmentation in the filamentary network of the Large Scale Structure, often referred to as "beads-on-a-string." This issue is particularly apparent in cosmologies with a suppression of the matter power spectrum at small spatial scales, such as warm dark matter models, where the perturbations induced by the N-body discretisation dominate the cosmological power at the suppressed scales. Initial conditions based on third-order Lagrangian perturbation theory, which allow for a late-starting redshift, have been shown to minimise numerical errors contributing to such artefacts. In this work, we investigate whether the additional use of a spatially adaptive softening for dark matter particles, based on the gravitational tidal field, can reduce the severity of artificial fragmentation. Tidal adaptive softening significantly improves force accuracy in idealised filamentary collapse simulations over a fixed softening approach. However, it does not substantially reduce spurious haloes in cosmological simulations when paired with such optimised initial conditions. Nevertheless, tidal adaptive softening induces a shift in halo formation times in warm dark matter simulations compared to a fixed softening counterpart, an effect not seen in cold dark matter simulations. Furthermore, initialising the initial conditions at an earlier redshift generally results in $z=0$ haloes forming from Lagrangian volumes with lower average sphericity. This sphericity difference could impact post-processing algorithms identifying spurious objects based on Lagrangian volume morphology. We propose potential strategies for reducing spurious haloes without abandoning current N-body methods.

\end{abstract}


%%%%%%%%%%%%%%%%% BODY OF PAPER %%%%%%%%%%%%%%%%%%

\section{Introduction} \label{sec:introduction}
The favoured paradigm of cosmology, $\Lambda$ cold dark matter ($\Lambda$CDM), describes the growth of structures through the continuous hierarchical merging of lower mass dark matter haloes into more massive ones, in an accelerating expanding Universe \citep{white_core_1978}. Cosmological N-body simulations have shown that one of the fundamental predictions of cold dark matter cosmologies is the abundance and distribution of small-scale structure that collapses gravitationally \citep[e.g.][]{springel_large-scale_2006, power_seeking_2013}. Such small-scale abundances distinguish between alternative dark matter models \citep[e.g.][]{bode_halo_2001, vogelsberger_subhaloes_2012, lovell_properties_2014, hui_ultralight_2017, tulin_dark_2018, nadler_constraints_2021}. Historically, cold dark matter cosmological simulations have shown an apparent overproduction of small-scale structure, clashing with observations of Milky Way satellites \citep{klypin_where_1999, moore_dark_1999}. However, advancements in baryonic models and improved computing power for developing state-of-the-art cosmological simulations have since helped align many $\Lambda$CDM  predictions with observations \citep[see][for a detailed review of unresolved tensions]{bullock_small-scale_2017, sales_baryonic_2022}. Despite this progress, predictions from a variety of different dark matter models remain valuable. Consequently,  significant efforts are being made to address the degeneracy between different dark matter cosmologies in the context of baryonic physics \citep[e.g.][]{lovell_addressing_2017, khimey_degeneracies_2021, alfano_breaking_2024, ussing_using_2024}.

Simulating alternative dark matter models is not without problems. Traditional N-body methods discretise the cosmic matter density field into phase-space particle tracers to follow the hierarchical growth of the small primordial density field fluctuations \citep[e.g.][]{efstathiou_numerical_1985, davis_evolution_1985}. Such discretisation leads to localised perturbations in the gravitational forces the particle tracers experience along their evolution \citep{binney_discreteness_2004}. When the force softening is chosen to be a fraction of the mean inter-particle distance, as it is often desired to balance two-body scattering noise and force bias \citep[e.g.][]{knebe_effects_2000, power_inner_2003, zhan_optimal_2006, price_energy-conserving_2007, romeo_discreteness_2008, zhang_optimal_2019, ludlow_numerical_2019}, localised perturbations lead to artificial fragmentation and the formation of spurious haloes. 

Simulations with suppressed small-scale power, such as warm dark matter (WDM) cosmologies, are strikingly affected by artificial fragmentation. In these simulations, perturbations dominate over physical velocities at the suppressed scales, causing filament fragmentation and leading to the formation of evenly spaced spurious haloes within the filamentary network \citep[often referred to as "beads-on-a-string",][]{wang_discreteness_2007, lovell_properties_2014}. However, \cite{power_spurious_2016} showed that artificial fragmentation is, in fact, a generic problem of N-body simulations, as the underlying collisionless assumption in the evolution of such systems fails when the amount of phase-space tracers is finite.

Various strategies have been adopted to address the growth of spurious haloes, including disregarding structure below a limiting mass scale \citep[e.g.][]{shen_thesan-hr_2024}, cleaning the simulations of spurious haloes in post-processing \citep[e.g.][]{lovell_properties_2014}, improving force calculations with spatially adaptive softening \citep[e.g.][]{hobbs_novel_2016}, and developing techniques beyond N-body methods that follow the evolution of dark matter in phase-space \citep[e.g.][]{angulo_warm_2013, sousbie_coldice_2016, hahn_adaptively_2016, stucker_simulating_2020, stucker_simulating_2022}. To this day, only the last approach has successfully eliminated the formation of spurious halos, provided that other sources of numerical errors (e.g. initial conditions, force calculations, or time-step determination) are also addressed. However, the tracking the phase-space dynamics of dark matter into the highly non-linear regime is algorithmically complex, and so far these techniques have been limited in their applications.

Inaccuracies in the initial conditions of cosmological simulations can also contribute to systematic errors in the non-linear evolution of Large Scale Structure. Contrary to common wisdom, \cite{michaux_accurate_2021} have shown that generating initial conditions at a later time using a higher and more computationally intensive order in Lagrangian perturbation theory (LPT) is preferable to starting earlier at a lower order. When the N-body system is initialised at earlier times, density perturbations are relatively small, and the self-interactions from the particle lattice tend to drive the N-body system's evolution toward a discrete configuration \citep{marcos_linear_2006, joyce_quantification_2007-1, joyce_quantification_2007, garrison_improving_2016}. Consequently, the earlier the starting time, the more the evolution of the discretised phase-space tracers will deviate from the expected fluid evolution. On the other hand, starting later at a higher LPT minimises the transients that arise due to truncations in the LPT order and reduces the spurious decaying modes that propagate from the linear regime to the non-linear evolution, hence producing the most accurate results to the smallest possible scales, using the least computational power to achieve it. However, on scales smaller than quasi-linear scales, i.e. those dominated by halo profiles, \cite{michaux_accurate_2021} note that initial conditions play a weaker role. For properties such as the abundance of collapsed structures measured by the halo mass function, especially at the low-mass end where artificial fragmentation is most apparent, their convergence also depends on further variables like the force resolution and time-stepping of the simulation \citep{reed_towards_2013, ludlow_numerical_2019, mansfield_how_2021}.

\cite{hobbs_novel_2016} revealed that anisotropic force errors, which are notably significant during the initial anisotropic collapse of filaments, play a major role in the formation of spurious halos. Their study examined various strategies for optimising gravitational softening to reduce two-body effects, minimise absolute force errors, and decrease the variation between different N-body realisations. To address these challenges and effectively manage the problems caused by force anisotropies, they concluded that force softening must be both spatially and temporally adaptive.

In this work, building on these insights, we aim to leverage the advantages of initialising simulations at lower redshift with higher-order LPT to minimise systematic discretisation errors and pair these optimised initial conditions with a novel tidal-based spatially adaptive force softening \citep{hopkins_novel_2023}. Tidal adaptive softening adjusts the force softening length according to the local tidal tensor, effectively capturing local gravitational anisotropy and scaling with local density. Moreover, it is designed to ensure that the two-body scattering noise remain subdominant to the smooth background gravitational forces. This feature is particularly relevant in cosmologies with suppressed small-scale power, where the intrinsic physical perturbations are weak, and numerical perturbations from discrete particle interactions could otherwise dominate and drive artificial fragmentation.

Individually, late-starting higher-order LPT initial conditions have been shown to reduce numerical initialisation errors, while adaptive softening techniques lead to more accurate calculations of forces. We aim to explore whether combining these methods can minimise the formation of artificial fragmentation in simulations.

This paper is organised as follows. In \Sec{sec:pipeline} we describe the numerical tools we use to create, evolve, and study the simulations for the analysis. In \Sec{sec:results} we quantify the performance of tidal adaptive softening on a simple model (\Sec{sec:results-subsec:asppwc}) and study its effect on more realistic cosmological simulations (\Sec{sec:results-subsec:cosmo}). In particular, in \Sec{sec:results-subsec:cosmo-subsubsec:chmf} we analyse the counts of collapsed structures in our set of cosmological simulations. In \Sec{sec:results-subsec:cosmo-subsubsec:shapes} we aim to quantify the influence of spurious structures on these counts, and in \Sec{sec:results-subsec:cosmo-subsubsec:zini} we focus on how the initial redshift of our initial conditions can influence the diagnostics used to identify spurious structures. Finally, we summarise our results in \Sec{sec:conclusions}. Throughout this analysis, unless stated otherwise, we assume a flat Planck18 cosmology \citep{planck_collaboration_planck_2020}, with $H_0=67.74$, $\Omega_{\rm M}=0.31$, $\Omega_{\Lambda}=0.69$, $\Omega_{\rm b}=0.048$, $n_{\rm s}=0.968$, and $\sigma_{8}=0.81$.

\section{Simulation pipeline} \label{sec:pipeline}
\subsection{Initial conditions}\label{sec:numerics-subsec:ics}
When required for our analysis, the initial conditions for the cosmological simulations are created with \monofonic{}\footnote{https://bitbucket.org/ohahn/monofonic/src/master/} \citep{michaux_accurate_2021}. \monofonic{} is a single resolution (mono-grid) distributed memory parallelised software package update of \music{} \citep{hahn_multi-scale_2011}, capable of generating initial conditions solutions up to third-order LPT (3LPT). To prevent aliasing arising from multiplying in real space quadratic and higher-order non-linearities, \monofonic{} temporarily enlarges the padding region of the computational domain in Fourier space, which allows aliased modes to be later discarded, preventing deviations in the evolution of physical wave modes \citep[i.e. following Orzag's 3/2 rule,][]{orszag_numerical_1971}. Additionally, to ensure that the particle ensemble follows the fluid evolution after the discretisation, \monofonic{} can compensate for the deviations by applying particle linear theory (PLT) corrections at early times when linear theory holds \citep{joyce_quantification_2007, marcos_particle_2008, garrison_improving_2016}.

%%%%%%%%%%%%%%%%%%%%%%%%%%%%%%

\subsection{Gravitational evolution}\label{sec:numerics-subsec:gravity}
The simulations in our analysis are evolved with the modular, multi-physics code \gizmo{}\footnote{http://www.tapir.caltech.edu/~phopkins/Site/GIZMO.html} \citep{hopkins_new_2015, hopkins_new_2017}. While \gizmo{} contains numerous modules for hydrodynamics and baryonic physics, as we are mainly interested in the artificial fragmentation arising in N-body methods, we focus our analysis on dark-matter-only simulations and avoid the extra layers of complexity that baryonic physics adds to our goal. 

The N-body gravity solver in \gizmo{} is an improved version of the \gadgetIII{} Tree-Particle Mesh hybrid (TreePM) scheme \citep{springel_cosmological_2005}. Among other common schemes, \gizmo{} includes a novel implementation of spatially adaptive force softening (and adaptive time-stepping) scaling with the tidal tensor of the gravitational potential \citep{hopkins_novel_2023}. For each particle $a$, their corresponding softening length $\epsilon_{a}$ varies as:
\begin{equation}\label{eq:tidal_softening}
    \epsilon_{a}=\xi\left(G m_{a} / \lVert\mathbf{T}\rVert_a \right)^{1/3},
\end{equation}
where $\xi\sim 2$ is a normalisation constant, $G$ is the Newtonian gravitational constant, $m_a$ the mass of particle, $\mathbf{T}=-\nabla\otimes\nabla\hat{\Phi}$ is the estimator of the tidal tensor corresponding to the local Newtonian potential estimator $\hat{\Phi}_a\equiv-(1/2) \sum_b m_{b}\hat{\phi}_{ab}$ \footnote{As discussed in \citet{hopkins_novel_2023}, the softening is defined in terms of an estimator of the local Keplerian potential $\hat{\phi}$ instead of the exact discrete potential used in calculating the gravitational forces. This is done because there may be situations in which it is numerically advantageous to use a different definition, and the choice of the softening rule is already somewhat arbitrary.}, and $\lVert\mathbf{T}\lVert^2=\sum\lambda_i^2$ is the Frobenius norm of the tidal tensor, defined as the sum of the squared eigenvalues $\lambda_i$ of the tidal tensor\footnote{Note that the trace of the tidal tensor of a particle Tr($\mathbf{T}_a$) $=-4 \pi G \rho_a$ scales with the local density. See \citet{hopkins_novel_2023} for a more in-depth discussion on the choice of gauge invariant quantities that can be computed from the tidal tensor $\mathbf{T}$ and their suitability as softening prescriptions.}.

Traditional local-neighbour adaptive force softening prescriptions, such as those scaling with the local density or particle number, are physically ill-defined when applied to a collisionless fluids. In the test particle limit where two nearby dynamically cold "cloud" particles approach each other, their collisionless nature dictates that they should "move through" each other, solely driven by the smooth, static, background gravitational field. However, as the cold particles approach each other, the local particle number (or density) increases. Therefore, their softening lengths shrink, deforming their respective spatial domains. Gravity-based spatially adaptive softening implementations naturally avoid this by instead evolving the softening based on the smooth gravitational background potential. Hence, the softening length minimally changes in such events \citep[for a more in-depth discussion and a comparison to other softening prescriptions, see][]{hopkins_fire-2_2018, hopkins_novel_2023}.

%%%%%%%%%%%%%%%%%%%%%%%%%%%%%%

\subsection{Structure identification}\label{sec:numerics-subsec:structureid}
The halo finding is performed using \ahf{}\footnote{http://popia.ft.uam.es/AHF/download.html} \citep{gill_evolution_2004, knollmann_ahf_2009}. To find potential halo centres, \ahf{} identifies local overdensities in an adaptively smoothed density field, automatically creating a hierarchical structure between haloes and its constituent substructure (subhaloes, sub-subhaloes, etc.). For every halo found, \ahf{} calculates their radius $r_{200}$ (and the corresponding enclosed mass $M_{200}$), determined as the radius $r$ at which the the density $\rho(r) = M(<r) / (4 \pi r^3 / 3)$ drops below $200\times\rho_{\rm crit}$, where $\rho_{\rm crit}$ the critical density of the Universe at a given redshift $z$. 

To trace the progenitors of haloes and subhaloes across snapshots we create merger trees with \mergertree{}, a tool packaged with \ahf{}. Each structure identified at redshift $z=0$ is traced back in time and assigned a main progenitor at each snapshot. The main progenitor is defined as the object that maximises the merit function $\mathcal{M}=N^2_{A\cap B}/(N_A N_B)$, where $N_A$ and $N_B$ are the particles in two haloes A and B (usually a descendent and a potential main progenitor), respectively, and $N_{A\cap B}$ is the number of particles found in both haloes. When progenitors of (sub)haloes are not found in a snapshot, the search continues in earlier snapshots (i.e. progenitors are allowed to "skip" snapshots), recovering an otherwise truncated branch of the merger tree \citep[e.g.][]{srisawat_sussing_2013, wang_sussing_2016}. Additionally, a limit on the mass ratio between a descendant and its potential main progenitor is applied afterwards to ensure the chosen main progenitor does not lead to an incorrect main tree branch. If the ratio limit is exceeded, the progenitor is discarded, and the halo search is continued in the following snapshot. For our analysis, we do not allow mass jumps between a descendant and its potential main progenitor greater than 2, i.e. M$_{\textrm{desc}}< 2$M$_{\textrm{prog}}$.

\section{Results} \label{sec:results}
\begin{figure*}
  \includegraphics[width=\textwidth]{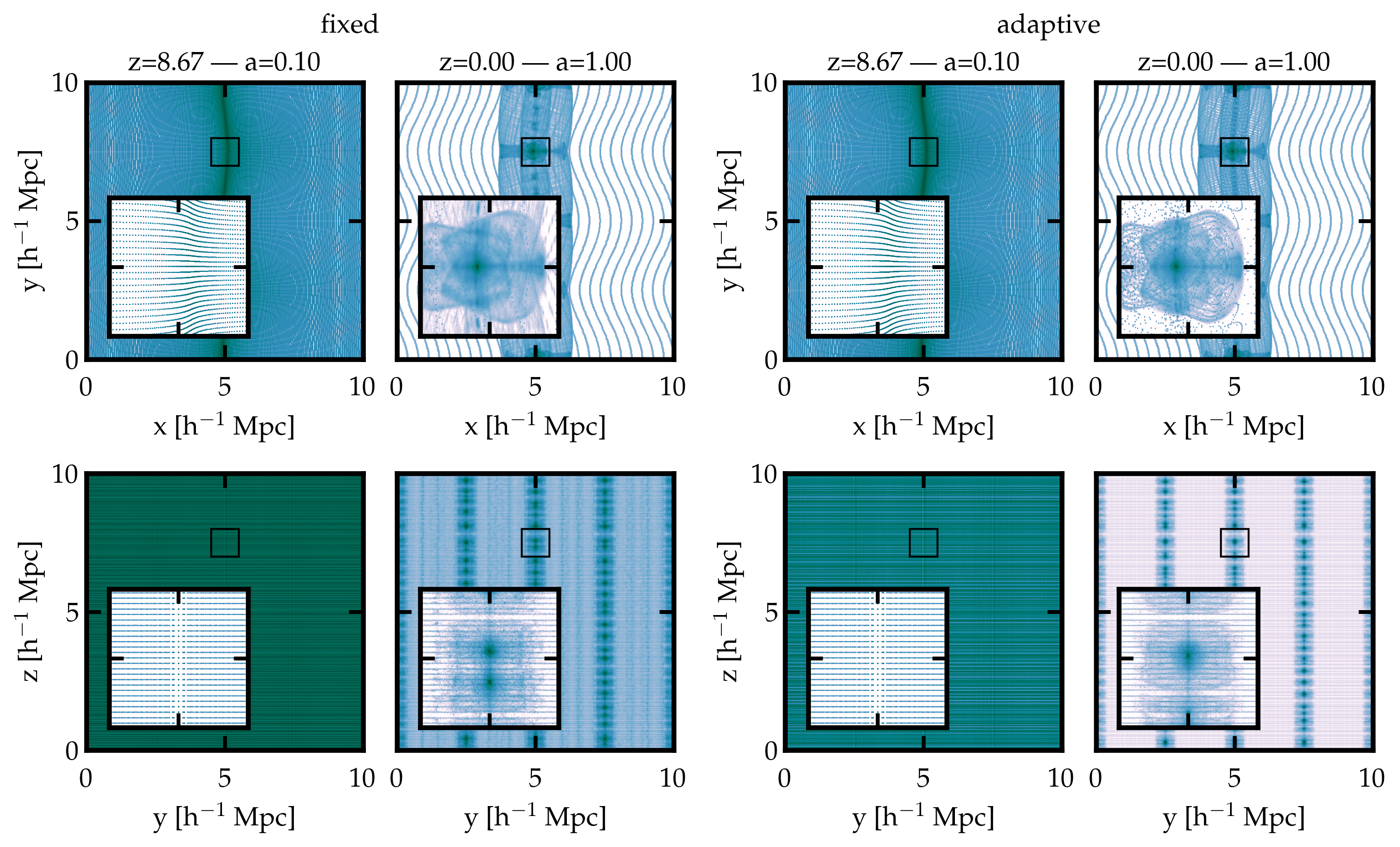}
  \caption{Projected densities of the anti-symmetrically perturbed plane wave collapse test for fixed (first and second columns) and tidal adaptive softening (third and fourth columns). For each softening type, we show the projected density at the shell crossing time ($a=a_c=0.1$) and at today's expansion factor ($a=1.0$). The top panels show the ($x,y$) projection, whereas the bottom panels show the ($y,z$) projection. In the inset we show a zoom of the square region highlighted in each sub-figure. The tidal adaptive softening run visibly leads to less artificial fragmentation along the $z-$axis and follows more accurately the evolution of the particles inside the filament compared the fixed softening run (see \Sec{sec:results-subsec:asppwc})}.\label{fig:density_compare}
\end{figure*}

Our analysis aims to integrate the recently developed enhancements in cosmological initial conditions with a new approach to adaptive softening to determine if artificial fragmentation can be prevented or mitigated in cosmological N-body simulations. Thus, it is instructive to examine first a simplified and well-established model of filamentary structure formation to determine the behaviour of tidal adaptive softening before proceeding to more accurate and complex cosmological simulations.

\subsection{Anti-symmetrically perturbed plane wave collapse}\label{sec:results-subsec:asppwc}

We investigate the performance of \gizmo{}'s tidal adaptive softening implementation using a simplified setup to measure spurious two-body effects that may arise from the discretisation of the cosmic matter density field in N-body simulations. This controlled environment allows us to isolate the role of the gravitational force softening scheme and quantify deviations from the expected physical evolution, which serve as a proxy for force computation errors that can seed spurious structures.

In this setup, we assume an $\Omega_{\textrm{M}} = 1, \Omega_\Lambda = 0$ Einstein-de Sitter cosmology. This is chosen to isolate the dynamics of gravitational collapse from the effects of accelerated cosmic expansion. The simulation uses a box side length of $L=10$ \hMpc{} containing $N_{\rm part}^3=256^3$ particles, a Hubble parameter $h= 0.7$, and a starting redshift $z_{\rm ini} = 99$ ($a=0.01$). The initial conditions of the simulations are set to follow the anti-symmetrically perturbed plane wave collapse described in \cite{valinia_gravitational_1997} and \cite{hahn_new_2013}, which model the formation of a homogeneous filament. In this test, a Zeldovich pancake plane wave is set in the x-direction with a phase perturbation in the y-direction, resulting in the gravitational collapse of a homogeneous filament along the z-direction. Due to the symmetry of this setup, no collapse should occur along the z-direction and any deviation from the initial z-positions of the particles will be
a result of artificial collapse. The initial gravitational potential of the filament is described by:
\begin{equation}\label{eq:valinia_perturbation}
    \phi(\mathbf{x}) = \bar{\phi} \cos \left( k_p \left[ x + \delta_a \frac{k_p}{k_a^2} \cos{k_a y} \right] \right),
\end{equation}
where $\mathbf{x}=(x,y,z)$, $\bar{\phi}$ is chosen so that the shell crossing scale factor $a_c$ happens at $a_c = 0.1$, $\delta_a=0.2$ is the amplitude of the phase perturbation, and $k_p =2\pi / L$ and $k_a = 4\pi / L$, are the pancake plane-wave and the anti-symmetric perturbation wave vectors, respectively. To determine the initial positions and velocities of the particles, we apply the Zeldovich approximation \citep{zeldovich_gravitational_1970} to an unperturbed Cartesian lattice. The initial conditions are then evolved to $z=0$ using both fixed softening and tidal adaptive softening in TreePM mode, with a $512^3$ PM mesh. The gravitational interaction of a particle is softened using a cubic spline kernel and becomes Newtonian at radial distances $r\geq 2.8\epsilon$, where $\epsilon$ is the Plummer-equivalent softening parameter controlling the force resolution of the simulation. Following \cite{power_inner_2003}, we set the Plummer-equivalent force softening for the fixed run to $1/40$ of the mean inter-particle distance at the initial conditions, i.e. $\epsilon_{\rm fixed}=977$ \hpc{} \footnote{The more recent analysis of \cite{ludlow_numerical_2019} revised the choice of optimal softening in cosmological simulations and concluded that, based on their study of convergence in central mass profiles of DM haloes, softening can be further lowered to $\sim 1/60$ of the mean inter-particle distance.}. For the adaptive run, we follow \gizmo{}'s documentation and set the minimum allowed softening to the smallest recommended softening length we want to allow in the simulation, determined by the fixed softening fiducial value, i.e. $1/10$ of the fixed softening, down to $\epsilon_{\rm adaptive}=97.7$ \hpc{}. With this procedure, we obtain two identical simulations, where the only difference is the force softening scheme used to evolve the formation of the homogeneous filament.

In \Fig{fig:density_compare}, we present the projected densities (i.e. $x-y$ and $y-z$) for the fixed and tidal adaptive softening simulations at the shell crossing time (at an expansion factor $a=a_c=0.1$) and at today’s expansion factor ($a=1.0$). At  $a=1.0$, both simulations show that particles have gravitationally collapsed along the $z-$axis, a result of artificial fragmentation after the formation of the filament. However, qualitatively, the tidal adaptive softening simulation more accurately captures the particle evolution after the shell crossing time. In the $x-y$ projections of the tidal adaptive softening, the inner structure of the filament is better resolved compared to the fixed softening simulation, where the particle positions appear more smeared out. This difference is even more pronounced in the $y-z$ projection. In the $z-$direction of the tidal adaptive runs, we observe significantly fewer particle clumps, which are shown as "banding" along the $z$-axis.

\begin{figure}
    \includegraphics[width=\columnwidth]{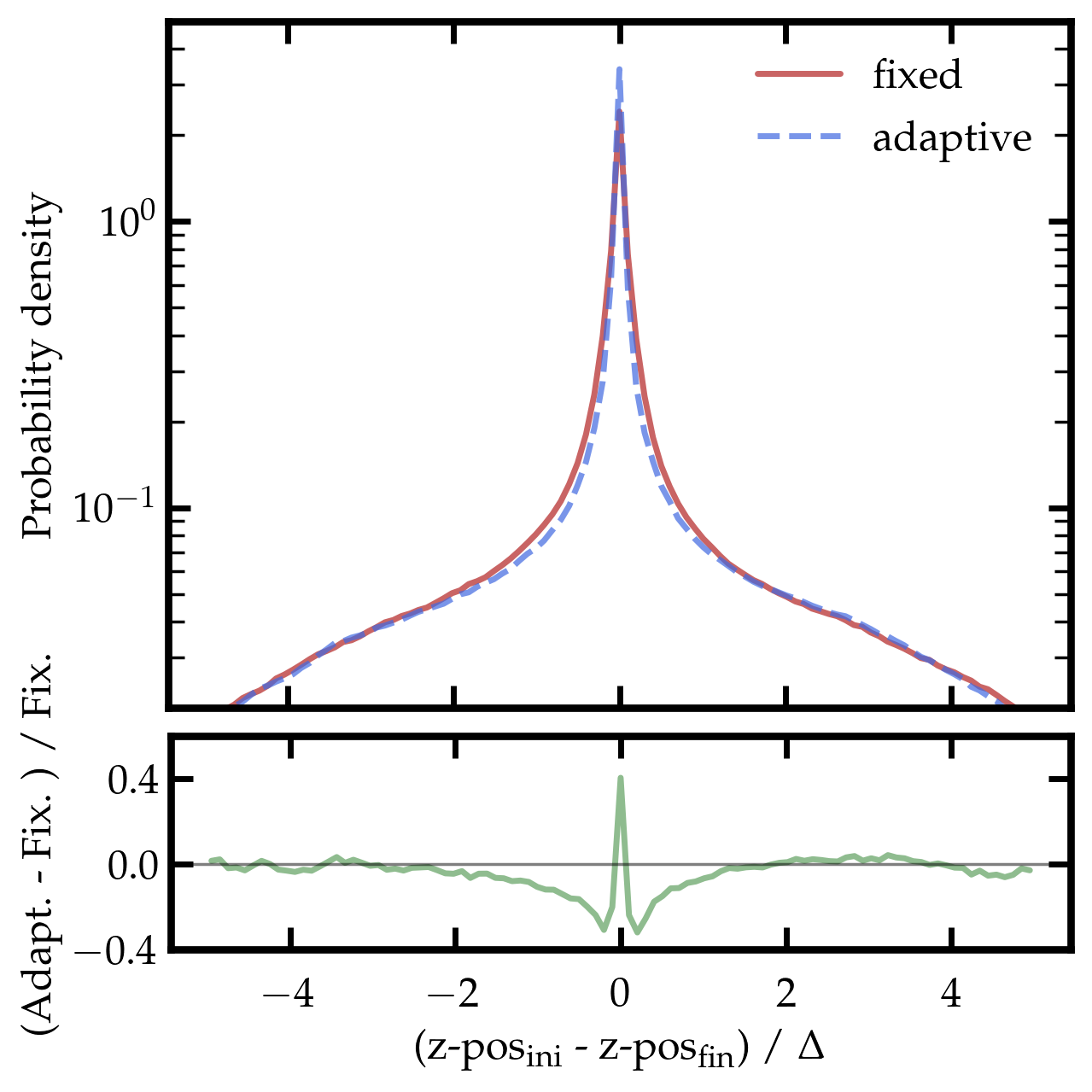}
    \caption{$z-$axis deviation from the initial uniform lattice of the particles in the anti-symmetrically perturbed plane wave collapse test. The top panel shows the probability density distribution of the difference between the $z-$axis position at the crossing time $a_c=0.1$ and the final position at $a=1.0$, normalised by the lattice spacing of the initial conditions. The bottom panel quantifies the relative difference between the shift probability distribution observed in the tidal adaptive softening with respect to the fixed softening. The horizontal line corresponds to no relative difference between the distributions. More particles in the tidal adaptive softening simulation retain their initial lattice position compared to the particles in the fixed softening counterpart run.}
    \label{fig:zshift_compare}
\end{figure}

To quantify the deviations from the expected evolution of the filament, we compare the positions of particles at crossing time $a_c=0.1$ and at $a=1.0$. We are particularly interested in the particle positions along the $z-$axis, as no physical perturbation is added along this axis in the initial conditions. Thus, particles should not experience any physical displacement from the initial lattice.

In \Fig{fig:zshift_compare} we show the probability density distribution of the difference between the $z-$axis position of the particles at crossing time $a_c=0.1$ and their positions at $a=1.0$, normalised by the inter-particle distance in the initial conditions lattice $\Delta$ (i.e. $\Delta=L/N_{\rm part}\simeq 39$ \hkpc{}). In the absence of force resolution errors, particles would not deviate from their initial positions, and their probability distribution would form a Dirac $\delta$ function centred along 0. While most particles show no deviation in both fixed and adaptive runs, both distributions exhibit $z-$position shifts more than 4 times the inter-particle distance from their initial positions. 

To better differentiate between fixed and tidal adaptive softening, we compare the relative difference between the distributions in the bottom panel, using the fixed softening run as our fiducial model. If both runs had similar particle shifts from their initial positions, there would be a minimal relative difference, and as such, the force resolution gained by using tidal adaptive softening would be minimal. However, compared with the fixed softening run, we identify 40 per cent more particles retaining their initial positions in the tidal adaptive softening run. These particles are located up to 2 lattice units (i.e. $\simeq 78$ \hkpc{}) along the $z-$axis away from their initial position in the fixed softening run. Therefore, tidal adaptive softening follows the evolution of the filament more accurately than an optimal fixed softening counterpart, even when the adaptive force softening is allowed to get 10 times smaller.

%%%%%%%%%%%%%%%%%%%%%%%%%%%%%%

\subsection{Cosmological simulations}\label{sec:results-subsec:cosmo}

\begin{figure*}
  \includegraphics[width=0.65\textwidth]{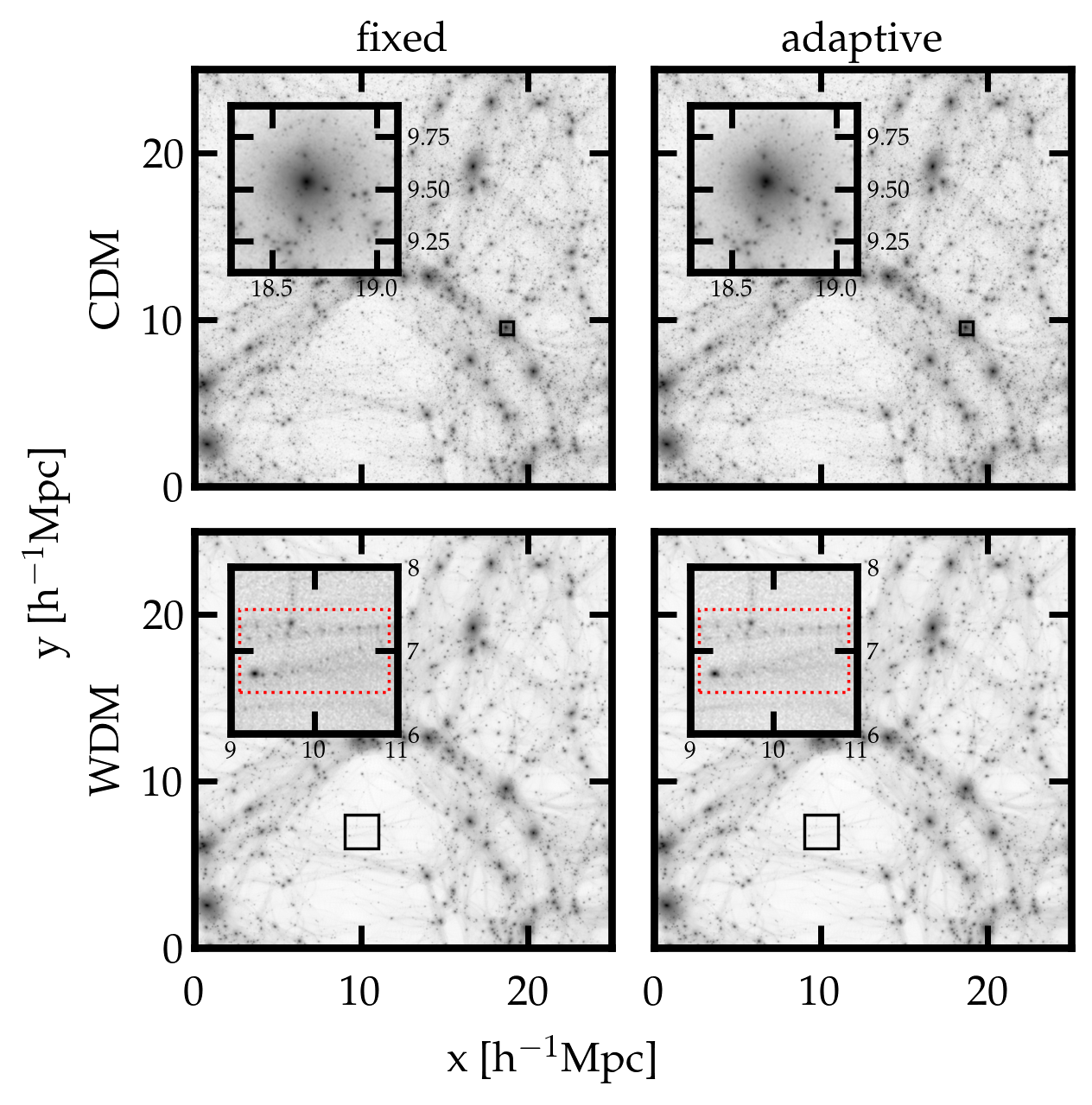}
  \caption{Projected densities of the cosmological simulation volumes at $z=0$. The top (bottom) panels corresponds to the cold (warm) dark matter simulations, whereas the left (right) columns shows the fixed (tidal adaptive) softening runs. The inset shows a zoom of the square region highlighted in each sub-figure. The filaments of the warm dark matter simulations show signs of artificial fragmentation in both the fixed and the tidal adaptive softening runs, as highlighted in the bottom panels by the red dotted rectangles. }\label{fig:proj_compare}
\end{figure*}

After characterising the performance of tidal adaptive softening in a simplified setup, we aim to investigate whether the improved modelling of the homogeneous filament, achieved through the use of tidal adaptive softening, translates to more accurate filamentary structure in cosmological simulations. This allows us to assess the combined effect of our choice of initial conditions and the tidal adaptive softening scheme on the subsequent abundance of spurious haloes formed via artificial fragmentation. To this end, we create a cold dark matter simulation and compare it with a warm dark matter counterpart, evolving both simulations with fixed and tidal adaptive softening. We then examine the extent of artificial fragmentation that arises in the warm dark matter simulations, as a result of the suppression of small-scale power in their initial conditions \footnote{While both cold and warm dark matter simulations suffer from artificial fragmentation, its effects are more conspicuous in warm dark matter cosmologies, as discussed in \cite{power_spurious_2016}.}, by analysing the abundance of haloes formed in each of the simulations, and characterising the contribution from spurious haloes to such abundances.

The initial conditions are set up at 3LPT using anti-aliasing and PLT corrections (see \Sec{sec:numerics-subsec:ics}), and at a starting redshift of $z_{\textrm{ini}}=39$. This is chosen to minimise the initialisation errors described in \cite{michaux_accurate_2021}. We choose a Cartesian lattice with a side length of $L=25$ \hMpc{} and $N_{\rm part}^3 = 512^3$ particles to model the cosmic dark matter density phase-space evolution. This corresponds to a  particle mass resolution of $1\times 10^7$ \hMsun{} (For convergence purposes, a lower resolution set of simulations has also been studied in \App{sec:appendix_a}). 

We calculate the cold dark matter transfer function using the \cite{eisenstein_baryonic_1998} fitting function. For the warm dark matter power spectrum, we subsequently truncate the cold dark matter power spectrum by applying the \cite{viel_constraining_2005} parametrisation of the transfer function corresponding to a warm dark matter particle candidate with a 1 keV mass. While such a candidate is in strong tension with observational constraints \citep[e.g.][]{nadler_constraints_2021, dekker_warm_2022, nadler_forecasts_2024},  our choice of a lighter mass ensures that the effects of artificial fragmentation are easily identifiable in the simulation.

Subsequently, we evolve the initial conditions to $z=0$ in 57 snapshots with both the fixed and the tidal adaptive softening implementations in \gizmo{}, using its TreePM mode with a $1024^3$ PM mesh and a cubic spline force kernel. Following the analysis in \Sec{sec:results-subsec:asppwc}, we choose a Plummer-equivalent force softening for the fixed run of $1/40$ of the mean inter-particle distance at the initial conditions, i.e. $\epsilon_{\rm fixed}=1.22$ \hkpc{}; and we set the minimum allowed tidal adaptive force softening to $1/10$ of the fixed force softening, i.e. $\epsilon_{\rm adaptive}=122$ \hpc{}. For the structure identification, we set \ahf{}'s minimum particle number threshold to 20 dark matter particles, which corresponds to a halo mass resolution to $2\times 10^{8}$ \hMsun{}.

In \Fig{fig:proj_compare} we present the projected density of the cosmological volumes at $z=0$ for each combination of dark matter and softening models we consider in our analysis. A closer examination of the zoomed-in regions reveals some key differences. In the cold dark matter simulations, we observe a distinct spatial distribution of the collapsed structures within the nodes of the large-scale structure. More notably, the results from the warm dark matter simulations (shown in the insets of the bottom panel) indicate that both the fixed and adaptive runs display signs of artificial fragmentation. This is suggested by the evenly spaced "beads" within their filamentary structures, which are highlighted by the red dotted rectangles in each inset.

\begin{figure}
  \includegraphics[width=\columnwidth]{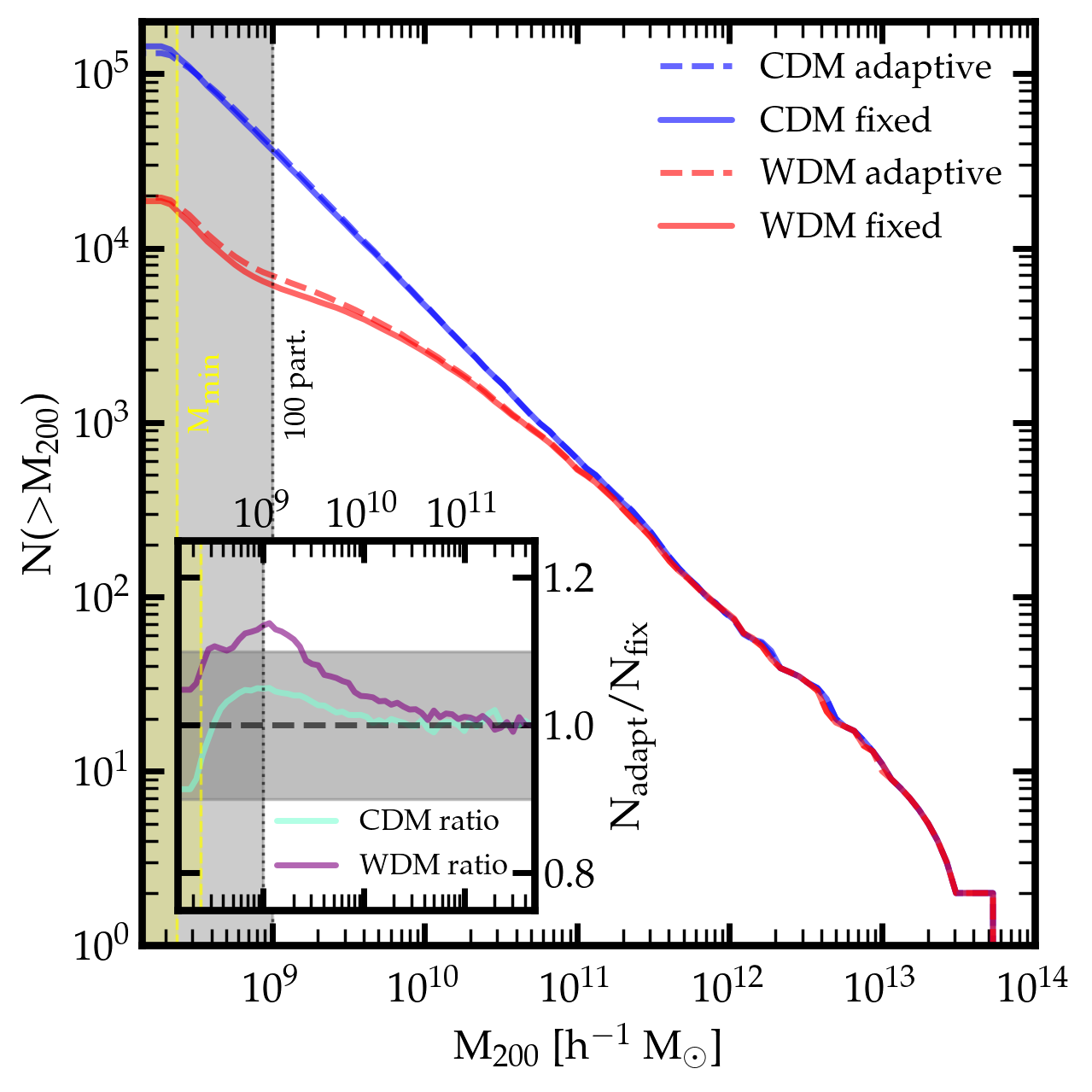}
  \caption{Cumulative halo mass function of the cold (blue) and warm (red) dark matter simulations at $z=0$, evolved with tidal adaptive (dashed) and fixed (solid) force softening. The vertical black dotted line marks the 100 particle halo limit, whereas the vertical yellow dashed line shows the \citet{lovell_properties_2014} minimum mass cut M$_{\textrm{min}}$ below which the mass function is dominated by spurious haloes. The inset shows the ratio of the tidal adaptive and the fixed softening cumulative halo mass functions for the cold (aquamarine) and warm (purple) runs. The shaded horizontal region shows the 20 per cent ratio deviation. Below the 100 particle limit, we identify the characteristic upturn in the cumulative halo mass function of the fixed and tidal adaptive softening warm dark matter runs caused by artificial fragmentation. }\label{fig:chmf_compare}
\end{figure}

%%%%%%%%%%%%%

\subsubsection{Cumulative halo mass functions}\label{sec:results-subsec:cosmo-subsubsec:chmf}

To quantify the differences in the collapsed structure between the simulations, in \Fig{fig:chmf_compare} we present the cumulative halo mass function at $z=0$ for the cold and warm dark matter simulations, comparing fixed and tidal adaptive force softening schemes. We observe good agreement among all runs above a halo mass of M$_{200}\sim10^{11}$ \hMsun{} ($\sim 10^4$ particles). At lower halo masses, the cumulative halo mass functions of the warm dark matter runs show a decrease in the number of identified haloes, as expected from the suppressed small-scale power initial conditions used for such simulations. Nonetheless, we identify a characteristic upturn in the warm dark matter mass function around M$_{200}=10^{9}$ \hMsun{} (corresponding to haloes with 100 particles) induced by artificial fragmentation \citep[e.g.][]{schneider_non-linear_2012, schneider_halo_2013, angulo_warm_2013, agarwal_structural_2015, hobbs_novel_2016}. 

Spurious fragmentation emerges at high redshift and triggers a subsequent gravitational collapse, leading to the formation of artificially-induced haloes. These spurious haloes then evolve into the non-linear regime at lower redshift, contributing to the count increase in the cumulative halo mass function below the suppressed scales at  $z=0$. \citet{lovell_properties_2014} showed that the proto-halos (the initial Lagrangian regions in the initial conditions) corresponding to spurious halos differ from those of physical halos, which we describe below. They proposed a mass cut M$_{\textrm{min}}$, based on a empirical mass limit M$_{\textrm{lim}}$ derived by \citet{wang_discreteness_2007}, below which the halo mass function would be dominated by spurious haloes. In our simulations, this mass cut M$_{\textrm{min}}\sim2.4\times10^8$ \hMsun{} (corresponding to haloes with $\sim$ 24 particles) lies below the 100 particle limit, i.e. at lower mass scales than the upturn in the cumulative halo mass function. We note this upturn is evident in both the fixed and adaptive runs of the warm dark matter cosmologies, with similar halo counts, indicating that the improved initial conditions and tidal adaptive softening does not eliminate such artificial fragmentation. 

We quantify the halo count differences between the tidal adaptive softening simulations and the fixed softening runs in the inset of \Fig{fig:chmf_compare}. By calculating the ratio of halo counts from the cumulative halo mass functions of the simulations, we observe a consistent deviation in the halo counts from the adaptive runs compared to their fixed softening counterparts. In the warm dark matter simulations, these differences are evident for masses below approximately $\sim10^{11}$ \hMsun{}, with a peak observed around $10^9$ \hMsun{} (near the 100 particle threshold). In this range, the adaptive run exhibits 14 per cent more halos than the fixed softening run. At lower masses, the halo count ratio returns towards unity, indicating that the observed excess is primarily concentrated around the 100 particle mass scale. Conversely, for cold dark matter simulations, the halo counts in both adaptive and fixed runs are more similar. Deviations begin to appear at lower masses than those seen in warm dark matter simulations (specifically around $\sim10^{10}$ \hMsun{}) yet peaking around similar halo masses as in the warm dark matter runs. We observe similar halo count ratio trends in the lower resolution runs described in \App{sec:appendix_a}. 

%%%%%%%%%%%%%

\subsubsection{Protohalo shapes}\label{sec:results-subsec:cosmo-subsubsec:shapes}

Next, we quantify the influence of spurious haloes in the observed halo counts in \Fig{fig:chmf_compare}. The analysis done by \citet{lovell_properties_2014} has shown that the shape of proto-haloes can be used to distinguish physical from artificial haloes: physical proto-haloes are spheroidal, while artificial ones tend to be flatter, oblate ellipsoids. Following their work, we select all the haloes and subhaloes identified at $z=0$ (i.e. contributing to \Fig{fig:chmf_compare}), and we determine if they are genuine or artificial haloes by analysing the sphericity of their proto-haloes at the initial conditions, i.e. $z_{\textrm{ini}}=39$. 

Subhaloes are prone to shape disruptions when they enter the environment of a host halo. To avoid biasing the shape analysis, we use the particle load of the progenitors of the $z=0$ sample at a time before any potential disruptions can strip away their constituent particles. Specifically, we select the earliest snapshot when a progenitor of a $z=0$ halo or subhalo reaches half of its maximum mass. Furthermore, to ensure accurate tracking of objects across snapshots, we only consider those identified in the simulations for at least 10 snapshots (allowing for snapshot skipping, as described in \Sec{sec:numerics-subsec:structureid}).

\begin{figure}
  \includegraphics[width=\columnwidth]{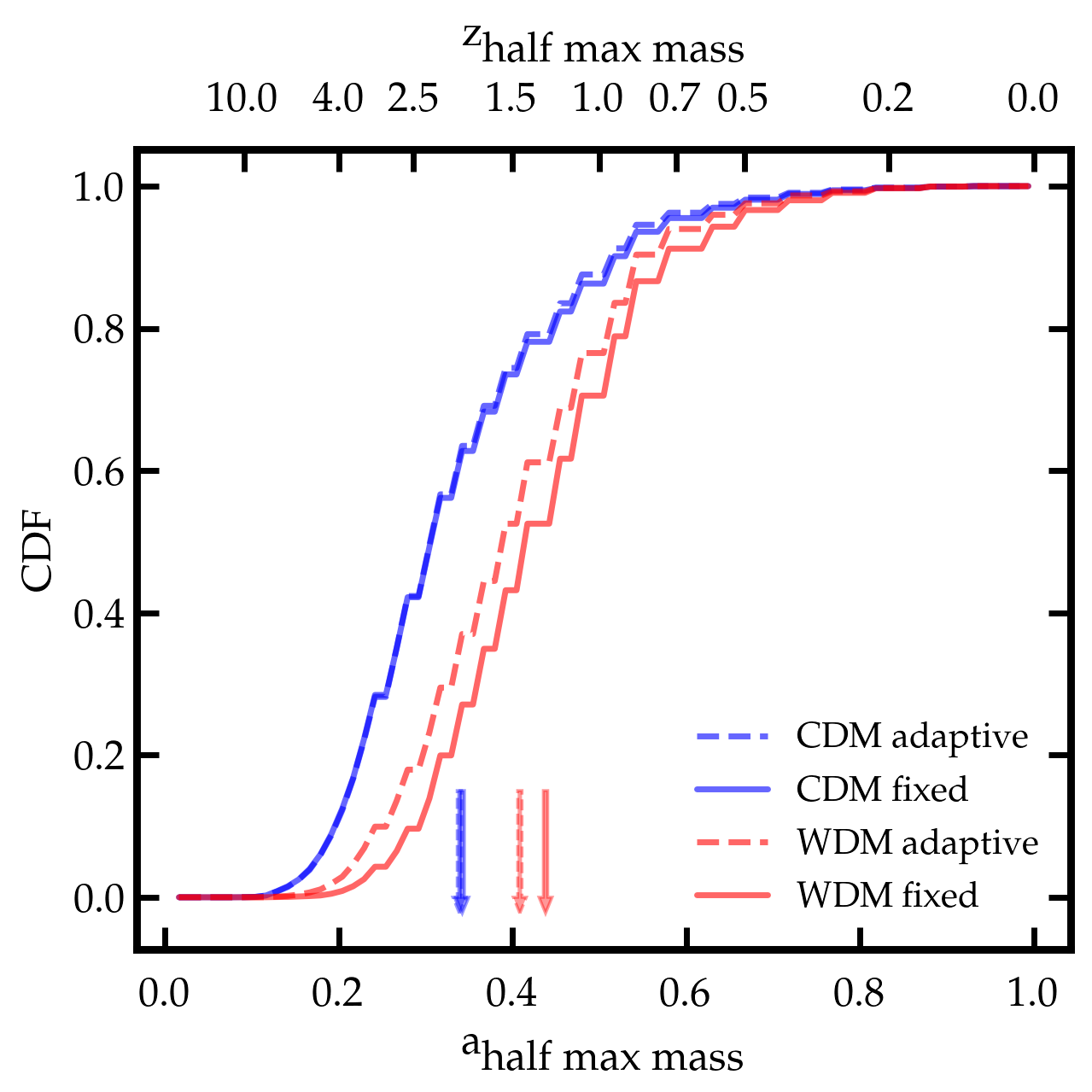}
  \caption{Normalised cumulative distribution function of the half-maximum mass scale factor (bottom axis) and redshift (top axis) of the haloes and subhaloes identified at $z=0$ in the fixed (dashed lines) and tidal adaptive softening (solid lines) runs of the cold (blue) and warm (red) dark matter cosmologies. The bottom arrows mark the corresponding mean values. The use of tidal adaptive softening and the presence of spurious haloes in the samples induce a shift in the cumulative distributions of the warm dark matter runs.}\label{fig:cdf_hmm}
\end{figure}

After applying the selection criteria mentioned above, we obtain samples of 107,890 and 101,743 haloes and subhaloes for the cold dark matter fixed and tidal adaptive softening runs, respectively. For the warm dark matter run, we obtain 10,452 objects for the fixed run and 13,060 for the tidal adaptive one. This represents 75 per cent and 78 per cent of the total number of halos and subhaloes identified at $z=0$ in the cold dark matter fixed and adaptive simulations, respectively, and 56 per cent and 67 per cent in the warm dark matter runs. In \App{sec:appendix_b}, we show the mass distribution of the samples in relation to the total number of objects identified at $z=0$.

We present the normalised cumulative distribution function (CDF) of the half-maximum mass scale factors (a$_{\textrm{hmm}}$) and redshifts for the four samples in our sphericity analysis in \Fig{fig:cdf_hmm}. The mean values of each distribution are indicated at the bottom of the figure. The distributions for the cold dark matter simulations are nearly identical, regardless of the softening method used (with mean values of $\bar{a}_{\textrm{hmm, fixed}}\sim$ $\bar{a}_{\textrm{hmm, adaptive}}=0.34$). However, in the warm dark matter simulations, we observe a difference of up to $\sim 0.1$ between the CDFs (at a$_{\textrm{hmm}}\sim 0.34$) resulting from the fixed softening and the tidal adaptive softening runs (with mean values of $\bar{a}_{\textrm{hmm, fixed}}=0.44$ and  $\bar{a}_{\textrm{hmm, adaptive}}=0.41$). This shift of $\sim 0.5$ Gyr in the mean of the warm dark matter half-maximum mass scale factor distributions, which could be induced by both the differing softening schemes and the potential inclusion of spurious haloes in the samples, is comparable to the delay in structure formation caused by varying the mass of the dark matter particle model, or equivalently, by adopting different cut-off scales in the power spectrum \citep[e.g.][where a similar shift was reported between the CDM and a 3 keV WDM model]{bose_copernicus_2016, nadler_cozmic_2024}. The relative contributions of the softening scheme and the presence of artificial fragmentation to this observed shift in WDM halo formation times will be explored further after we analyse the properties of these halo populations in more detail.

After identifying the progenitor and its half-maximum mass snapshot for each $z=0$ object in the four samples, we select the constituent particles of each progenitor halo at their respective half-maximum mass snapshot and trace these selected particles back to the initial conditions. These particle distributions within the initial conditions (i.e. the Lagrangian volume of a progenitor)  
define the proto-haloes corresponding to the $z=0$ samples. It is for these primordial proto-haloes, using their particle positions in the initial conditions at $z_{\textrm{ini}}=39$, that we determine their shape by calculating their inertia tensor:
\begin{equation}\label{eq:tensor_inertia}
    I_{ij} = \sum_{\textrm{particles}}{m\left(\delta_{ij}|\mathbf{x}|^2 - x_i x_j\right)},
\end{equation}
where $m$ is the dark matter particle mass, $\delta_{ij}$ is the Kronecker delta function, and $\mathbf{x}$ is the particle position with respect to the proto-halo's centre of mass at the initial conditions. Diagonalising the tensor of inertia we find the three axis lengths $\lambda_c\leq\lambda_b\leq\lambda_a$ of the proto-halo. The sphericity $s$ is, therefore, defined as the ratio of the shortest over the longest axis of the triaxial proto-halo in the initial conditions, i.e. $s=\lambda_c / \lambda_a$. According to this definition, a particle distribution with a sphericity close to 1 resembles a sphere, while values closer to 0 indicate an oblate ellipsoid.

\begin{figure}
  \includegraphics[width=\columnwidth]{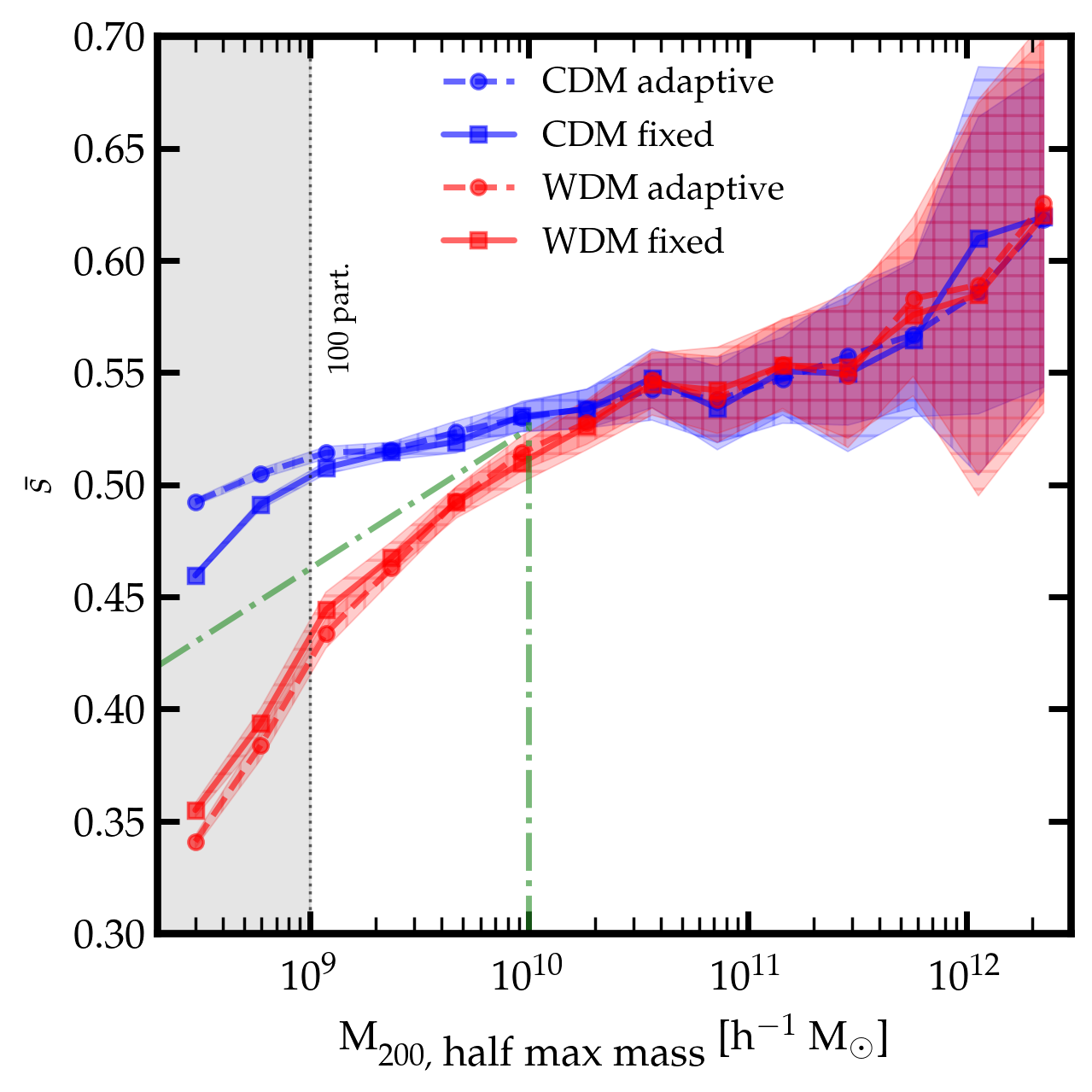}
  \caption{Mean sphericity $\bar{s}$, calculated for proto-haloes in the initial conditions, as a function of the half-maximum mass of the main progenitors of the $z=0$ objects. These proto-haloes correspond to the Lagrangian volumes of the haloes and subhaloes identified at $z=0$. In blue, cold dark matter simulations ran with tidal adaptive (dashed line with circle markers) and fixed (solid line with square markers) force softening; whereas in red, we show the warm dark matter counterpart simulations. The coloured hatched regions represent the 99 per cent bootstrapped confidence levels of each mean trend. We use horizontal hatches for the fixed runs and vertical hatches for the adaptive ones. The vertical black dotted line and shaded region marks the 100 particle limit. We show the empirical mass dependence of the mean sphericity of the samples with a green dash-dotted line. The vertical green dash-dotted line marks the mass threshold we use for the empirical mean sphericity relation. Objects with sphericities closer to 1 resemble a sphere, whereas sphericities closer to 0 indicate an oblate ellipsoid. Spurious haloes have lower sphericities than genuine ones.}\label{fig:sphericity_compare}
\end{figure}

In \Fig{fig:sphericity_compare} we show the mean sphericity of the four samples as a function of their corresponding progenitor half-maximum mass. We observe a good agreement among the simulations for haloes with half-maximum masses above M$_{200, \textrm{ hmm}}\sim2\times10^{10}$ \hMsun{}, with a slight decrease in sphericity of $\sim 0.1$ over two orders of magnitude in halo mass across all runs. However, as noted by \citet{lovell_properties_2014}, there is a more significant decrease in sphericity for warm dark matter simulations compared to their cold dark matter counterparts for masses M$_{200, \textrm{ hmm}}<2\times10^{10}$ \hMsun{}. This discrepancy is attributed to the influence of spurious haloes on the mean value\footnote{The presence of spurious haloes in the sample becomes apparent from the upturn in the halo mass function shown in \Fig{fig:chmf_compare}, which occurs below the 100 particle limit. However, the mean sphericity analysis illustrated in \Fig{fig:sphericity_compare} indicates that spurious haloes can significantly influence the trend even above this limit.}. The decrease in sphericity is observed in both fixed and tidal adaptive warm dark matter simulations, following similar trends. This suggests that tidal adaptive softening does not have a significant impact on the sphericity of artificial halos. However, at around the 100 particle limit, the mean sphericity trends of the cold dark matter fixed and tidal adaptive simulations begin to diverge, with differences reaching up to $\sim 0.05$ at M$_{200, \textrm{ hmm}} = 3 \times 10^8$ \hMsun{}.

We quantify number of spurious haloes in warm dark matter simulations by visually defining a simple empirical mean sphericity relation from the mean sphericity trend of the simulations in order to distinguish spurious haloes from genuine ones below the half-maximum mass threshold M$_{200, \textrm{ hmm}}=10^{10}$ \hMsun{}, where the sphericity trends of the cold and warm dark matter runs diverge:  
\begin{equation}\label{eq:medsph_mass_trend}
    \bar{s} = \alpha \log_{10}\left(\textrm{M}_{200, \textrm{ hmm}} / \hMsun{}\right) + \beta,
\end{equation}
with $\alpha=0.0624$ and $\beta=-0.0988$. These parameters were determined by defining a line that visually separates the mean proto-halo sphericity trends of the CDM and WDM samples in \Fig{fig:sphericity_compare} over the mass range where their sphericity values diverge, specifically from min(M$_{200,\textrm{ hmm}}) = 3\times 10^8\hMsun{}$ (with an adopted corresponding mean sphericity $\bar{s}_{\textrm{min}}=0.43$) to max(M$_{200, \textrm{ hmm}}) = 10^{10}\hMsun{}$ (with  
$\bar{s}_{\textrm{max}}=0.525$). The slope $\alpha$ and intercept $\beta$ were calculated to ensure the line passes through these two defined points. 

Using this empirical mean sphericity relation, we aim to estimate the level of artificial fragmentation in our simulations. Objects with a mass below the half-maximum threshold and sphericities lower than the empirical relation are classified as spurious.\footnote{Note that our half maximum mass-dependent empirical relation was visually derived by accounting for the mean sphericity trends of the simulations and their bootstrapped confidence levels. Thus, due to the overlap in the sphericity distributions, CDM haloes could also be classified as spurious. We focus on the WDM runs, as the effects of artificial disruption are more pronounced in these simulations.}.

\begin{table}
\centering
\begin{tabular}{@{}l|cl|cl|@{}}
\cmidrule(l){2-5}
\multirow{2}{*}{} & \multicolumn{2}{c|}{fixed} & \multicolumn{2}{c|}{adaptive} \\ \cmidrule(l){2-5} & \multicolumn{1}{c|}{total} & \multicolumn{1}{c|}{artificial} & \multicolumn{1}{c|}{total} & \multicolumn{1}{c|}{artificial} \\ \midrule

\multicolumn{1}{|l|}{WDM} & \multicolumn{1}{l|}{10,452} & 5,973 (57.1\%) & \multicolumn{1}{l|}{13,060} & 8,004 (61.3\%) \\ \bottomrule
\end{tabular}
\caption{Number of artificial haloes identified in the warm dark matter fixed and tidal adaptive runs based on the empirical mean sphericity mass relation in \Eq{eq:medsph_mass_trend}. In parenthesis we show the percentage of the total halo sample that is spurious.}\label{tab:arthalo_counts}
\end{table}

In \Tab{tab:arthalo_counts}, we present the total number of haloes and subhaloes in the warm dark matter samples, along with the count of those classified as artificial based on our empirical relationship. We find that 57.1 (61.3) per cent of the warm dark matter objects with a mass below M$_{200, \textrm{ hmm}}=10^{10}$ \hMsun{} are classified as artificial in the fixed (tidal adaptive) softening run. Varying the slope of the empirical relationship by 30 per cent, we observe count differences of up to 15 per cent. Overall, we find similar total counts regardless of the softening method and the parameters of the empirical relationship used, although the tidal adaptive softening consistently results in $\sim 4$ per cent more haloes compared to the fixed softening approach.

In light of these results, we re-examine the observed shift in the half-maximum mass scale factor distribution for WDM haloes seen in \Fig{fig:cdf_hmm} by attempting to minimise the contribution from haloes likely to be spurious. Based on the mean sphericity trends of \Fig{fig:sphericity_compare}, which indicate that WDM haloes with M$_{200, \textrm{ hmm}}>10^{10}$ \hMsun{} have proto-halo sphericities more consistent with genuine CDM structures, we reproduced the normalised CDF of $a_{\textrm{hmm}}$ for this more massive, likely genuine WDM halo sample.

This analysis reveals that even for this sub-sample of haloes with M$_{200, \textrm{ hmm}}>10^{10}$ \hMsun{}, a statistically significant difference in formation times between the fixed and tidal adaptive softening WDM runs persists, albeit reduced compared to the full sample (mean values of $\bar{a}_{\textrm{hmm, fixed}}=0.38$ and  $\bar{a}_{\textrm{hmm, adaptive}}=0.37$). This persistent, though smaller, shift suggests that while spurious haloes may contribute to the magnitude of the difference seen in the full sample in \Fig{fig:cdf_hmm}, the tidal adaptive softening scheme itself also appears to intrinsically influence the mass assembly history and hence the inferred formation times of genuine WDM haloes. 

Nevertheless, despite utilising a higher order of LPT and a later start in generating the initial conditions, as well as the improved force evolution modelling shown in the simplified setup in \Sec{sec:results-subsec:asppwc}, artificial fragmentation remains present in our warm dark matter cosmological simulations. Moreover, the use of tidal adaptive softening does not significantly reduce the number of spurious halos compared to the fixed softening scheme.

\begin{figure}
  \includegraphics[width=\columnwidth]{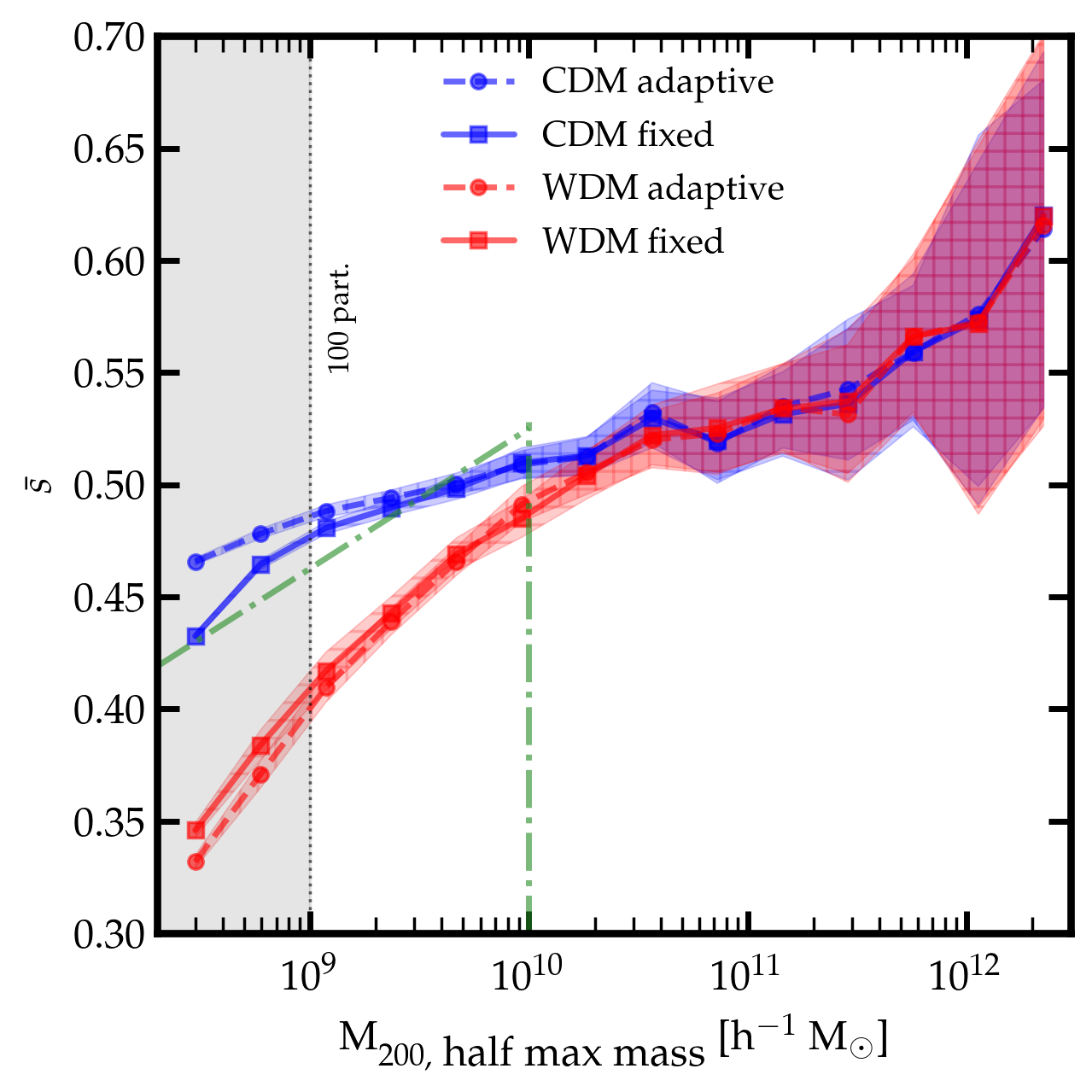}
  \caption{Same as \Fig{fig:sphericity_compare} but for runs initialised at redshift z$_{\textrm{ini}}=99$. An earlier starting redshift of the simulations leads to overall less spherical haloes and subhaloes.}\label{fig:sphericity_compare_z99}
\end{figure}

%%%%%%%%%%%%%

\subsubsection{Initial redshift influence on protohalo shapes}\label{sec:results-subsec:cosmo-subsubsec:zini}

We finalise our analysis by examining how the initial redshift of the initial conditions affects the sphericity results. We use the same setup for the initial conditions as in our previous analysis, but initialise them at $z_{\textrm{ini}}=99$, while maintaining the same LPT order and corrections we applied for our $z_{\textrm{ini}}=39$ simulations. We follow the same procedure outlined above and determine the mean sphericity of the corresponding four samples as a function of their half-maximum mass. The resulting trends are illustrated in \Fig{fig:sphericity_compare_z99}. 

Compared to the late-start simulations, the mean sphericity trends are fairly similar. We observe a split in the cold and warm dark matter runs at a similar half-maximum mass threshold of M$_{200}=10^{10}$ \hMsun{}. However, we find an overall decrease of sphericity of $\sim 0.03$ (6 per cent) compared to the $z_{\textrm{ini}}=39$ simulations. This effect is less pronounced in the lower mass bins of the warm dark matter simulations compared to the cold dark matter runs. The mean sphericity of the cold dark matter tidal adaptive softening run shows a deviation from the fixed softening counterpart at half-maximum masses above the 100 particle mass limit, at similar a mass range as in the $z_{\textrm{ini}}=39$ simulations. We note that the empirical sphericity relation we obtain from the $z_{\textrm{ini}}=39$ analysis (i.e. \Eq{eq:medsph_mass_trend}) is no longer separating the cold and warm dark matter mean sphericity trends in the $z_{\textrm{ini}}=99$ counterpart simulations, as the $z_{\textrm{ini}}=99$ mean sphericities are overall lower. Below the half-maximum mass threshold, a considerable number of cold dark matter objects fall below the empirical relation. This suggests that the initial redshift plays an important role in the post-processing of spurious haloes, as the shapes of the Lagrangian volumes corresponding to $z=0$ objects are sensitive to the time at which simulations are initialised.

\section{Conclusions} \label{sec:conclusions}
In this analysis, we investigate the formation of spurious halos in N-body simulations and test whether combining two novel enhancements, namely late-time accurate initial conditions and tidal adaptive softening, can reduce artificial fragmentation in these simulations.

We begin by examining the effects of tidal adaptive softening in a simplified setup of filamentary collapse. To do this, we create initial conditions that mimic the formation of a homogeneous filament and evolve them using both a fixed and a tidal adaptive force softening scheme. We quantify the performance of the tidal adaptive softening approach by comparing the degree of artificial fragmentation in the simulation to the fixed softening counterpart.

Next, we move on to more accurate filamentary structure in cosmological simulations. We generate initial conditions aimed at minimising initialisation errors commonly arising in cosmological runs using low orders of Lagrangian Perturbation Theory (LPT) at high initial redshift. These initial conditions are then evolved with both tidal adaptive and fixed softening methods. To facilitate the identification of spurious haloes, we generate cold and warm dark matter counterpart simulations and analyse the haloes and subhaloes found at $z=0$. By examining the sphericity of the proto-haloes of the objects identified at $z=0$, we quantify the amount of artificial haloes formed in each simulation. 

The main results of our analysis are the following:
\begin{itemize}
    \item In the simplified setup of the gravitational collapse of a homogeneous filament, particles are expected to collapse along the $x$ and $y-$axis yet retain their initial lattice position along the $z-$axis. Up to 40 per cent more particles evolved under tidal adaptive softening simulation retain their lattice position over an "optimal" fixed softening choice, while allowing for softening lengths to reach 10 times smaller values than in the fixed softening run.
    \item However, the force accuracy improvement observed in the homogeneous filament collapse does not directly carry over to cosmological simulations. The advantages gained from initialising the initial conditions at lower redshift and using higher-order LPT, combined with the benefits of evolving the simulations under a tidal adaptive softening scheme, do not mitigate the formation of spurious haloes in the simulations. When analysing the cumulative halo mass functions of warm dark matter simulations, we observe similar characteristic upturns below the suppressed mass scales (induced by artificial haloes) in both fixed and tidal adaptive softening runs.
    \item We quantify the number of spurious halos and subhaloes that contribute to the upturn in the cumulative halo mass function in the warm dark matter simulations. Based on an empirical relation for the mean sphericity visually derived from our analysis, we find that 57 per cent of the objects in the fixed softening run are artificial. In the tidal adaptive softening run, 61 per cent of the objects are considered spurious.
    \item Nonetheless, the effects of the improved initial conditions and tidal adaptive softening are not negligible. The tidal adaptive softening simulations consistently show a surplus of objects that peaks at around the 100 particle mass limit (M$_{200}=10^9$ \hMsun{}) when compared to simulations with fixed softening. This difference is especially pronounced in the warm dark matter simulations, where the tidal adaptive softening run can yield up to 1.14 times more objects than the fixed softening simulations.
    \item We analyse the half-maximum mass scale factor distributions of our samples, which is defined as the earliest snapshot at which an object reaches half of its maximum mass. By comparing these distributions, we observe a shift between the warm dark matter fixed and tidal adaptive softening runs of $\bar{a}_{\text{hmm, fixed}} - \bar{a}_{\text{hmm, adaptive}} = 0.03$ ($\sim 0.5$ Gyr). This shift is similar to the delay in structure formation caused by varying the mass of the warm dark matter candidate or, equivalently, the power spectrum cut-off in the initial conditions \citep[e.g.][where a similar shift was reported between the CDM and the 3 keV WDM simulations in their analysis]{nadler_cozmic_2024}. In contrast, no such shift is observed in the cold dark matter runs. 
    \item The shift in the half-maximum mass scale factor distributions of the WDM fixed and tidal adaptive softening runs is stiller present when only considering samples with little to no spurious haloes, albeit the shift is smaller than in the full sample. This suggests that the use of tidal adaptive softening is intrinsically influencing the inferred timing of halo mass assembly in the WDM runs. This indicates that caution is needed when interpreting structure formation predictions that involve varying WDM masses. Numerical issues appear to be at least as significant in these analyses.
    \item When comparing the mean sphericities of the simulations, we observe an additional trend divergence between the cold dark matter fixed and the tidal adaptive softening runs for half-maximum masses below the threshold of 100 particles. Specifically, at a half-maximum mass of M$_{200, \textrm{ hmm}} = 3 \times 10^8$ \hMsun{}, the mean sphericity of the tidal adaptive run is approximately 0.05 higher than that of the fixed run. This difference is not present in the WDM simulations.
    \item By initialising the initial conditions at a higher redshift (z$_{\textrm{ini}}=99$) and conducting the same analysis, we observe an overall decrease in the mean sphericity of our sample of $\sim 0.03$ (6 per cent) compared to the runs with z$_{\textrm{ini}}=39$. This indicates that the initial redshift at which a cosmological simulation is set can influence the post-processing algorithms that depend on the shape properties of proto-haloes.
    
\end{itemize}

Additional analysis is needed to understand the extent to which tidal adaptive softening and the initial conditions setup can influence the identification of spurious halo algorithms in cosmological simulations, as illustrated by the shift in the normalised cumulative half maximum mass scale factor distributions of the fixed and tidal adaptive softening warm dark matter simulations and the overall reduction in mean sphericity in simulations initialised at earlier redshifts.

Nevertheless, given our results, it is clear that the current implementation of tidal adaptive softening is not effective in reducing N-body artificial fragmentation. Therefore, new approaches should be explored. One possible solution is to change the definition of the tidal adaptive softening. The existing implementation in \gizmo{}, described by \Eq{eq:tidal_softening}, defines the softening scaling based on the relationship $\epsilon_a \propto 1/\lVert\mathbf{T}\rVert_a^{1/3} \propto 1/(\sum \lambda_i^2)^{1/6} \propto 1 / \textrm{max } \lambda_i^{1/3}$. In other words, the softening is dependent on the highest eigenvalue of the tidal tensor, which corresponds the direction of the strongest gravitational gradient.

An alternative approach worthy of future investigation could be to modify \gizmo{}’s tidal adaptive softening, for instance by considering timesteps based on the lowest eigenvalue of the tidal tensor of inertia instead. We note that this choice is somewhat arbitrary, and such a modification would shift the focus of the scenarios that \gizmo{} targets: the current implementation emphasises improved force resolution during violent mergers and the resolution of tidal streams, while the proposed change would enhance resilience against anisotropic scales and tidal gradients during filament collapse. Although this modification appears straightforward, a thorough analysis is necessary to ensure that it successfully compares with other alternative softening schemes, as outlined in \citet{hopkins_novel_2023}.

Our work highlights that even with the combined advancements of late-starting, higher-order initial conditions and a novel tidal adaptive softening scheme, the problem of artificial fragmentation persists in simulations with suppressed small-scale power. This highlights that reliably predicting the low-mass halo population in alternative dark matter scenarios remains a significant challenge for current N-body methods. Potential avenues for further improvement may lie in modifications to the adaptive softening criteria themselves, such as basing them on different aspects of the tidal tensor. Fully resolving such numerical challenges is crucial for obtaining robust astrophysical predictions needed to constrain the nature of dark matter.

\section*{Acknowledgements}
This research was supported in part by the Australian Government through the Australian Research Council Centre of Excellence for Dark Matter Particle Physics (CDM, CE200100008). DC acknowledges the support of an ARC Future Fellowship (FT220100841). This work was performed on the Ngarrgu Tindebeek/OzSTAR national facility at Swinburne University of Technology. The OzSTAR program receives funding in part from the Astronomy National Collaborative Research Infrastructure Strategy (NCRIS) allocation provided by the Australian Government, and from the Victorian Higher Education State Investment Fund (VHESIF) provided by the Victorian Government. CP acknowledges the support of the ARC Centre of Excellence for All Sky Astrophysics in 3 Dimensions (ASTRO 3D), through project number CE170100013.  AK is supported by the Ministerio de Ciencia e Innovaci\'{o}n (MICINN) under research grant PID2021-122603NB-C21 and further thanks H\"usker D\"u for new day rising. AU acknowledges the support of an Australian Government Research Training Program Scholarship.

We thank our anonymous referee for their thoughtful report and valuable suggestions. RMP thanks Jens St{\"u}cker for his valuable comments and for the help provided setting up initial conditions for the filamentary collapse, Phil Hopkins for valuable discussions and ideas during the project, and Ciaran O'Hare for sharing his \textsc{MATPLOTLIB} style file.

This work uses the \textsc{SCIPY} \citep{virtanen_scipy_2020}, \textsc{NUMPY} \citep{harris_array_2020}, \textsc{MATPLOTLIB} \citep{hunter_matplotlib_2007}, and \textsc{PANDAS} \citep{mckinney_data_2010}
packages for \textsc{PYTHON}. 

%%%%%%%%%%%%%%%%%%%%%%%%%%%%%%%%%%%%%%%%%%%%%%%%%%
\section*{Data Availability}
The simulations used in this study are available and accessible on reasonable request. The data products of this work can be shared upon request to the corresponding author.



\bibliographystyle{mnras}
\bibliography{references}

%%%%%%%%%%%%%%%%%%%%%%%%%%%%%%%%%%%%%%%%%%%%%%%%%%

%%%%%%%%%%%%%%%%% APPENDICES %%%%%%%%%%%%%%%%%%%%%

\appendix
\section{Resolution test}\label{sec:appendix_a}
\begin{figure}
  \includegraphics[width=\columnwidth]{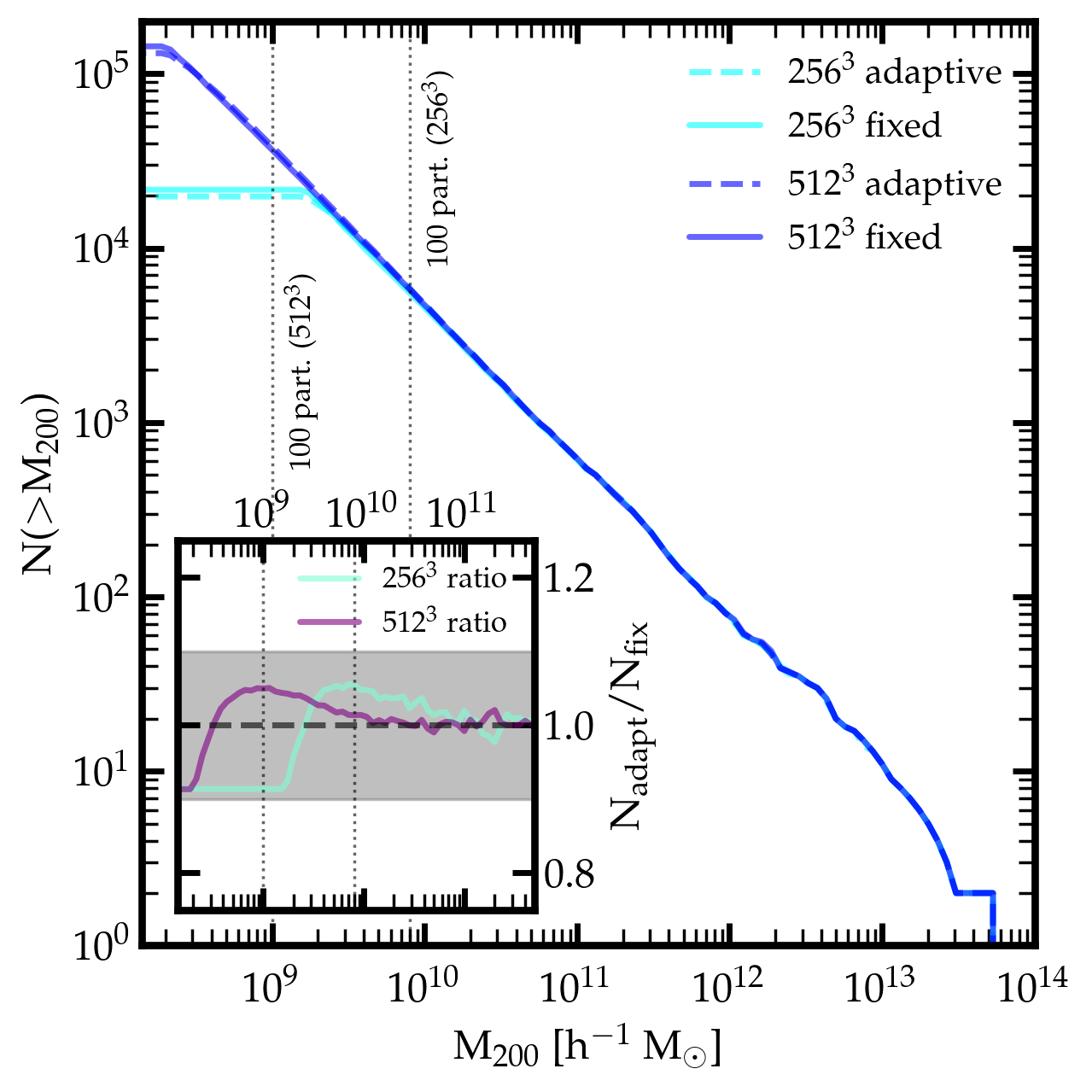}
  \caption{Cumulative halo mass function of the cold low resolution (N$_{\rm part}^3=256^3$, blue) and the cold at higher resolution (N$_{\rm part}^3=512^3$, cyan) dark matter simulations, evolved with tidal adaptive (dashed) and fixed (solid) force softening (similar to \Fig{fig:chmf_compare}). The vertical black dotted lines mark the 100 particle limit corresponding to each simulation. The inset shows the ratio of the tidal adaptive and the fixed softening cumulative halo mass functions for the low (aquamarine) and high (purple) resolution runs. The shaded region shows the 20 per cent ratio deviation.}\label{fig:a1_chmf_res_cdm_compare}
\end{figure}

\begin{figure}
  \includegraphics[width=\columnwidth]{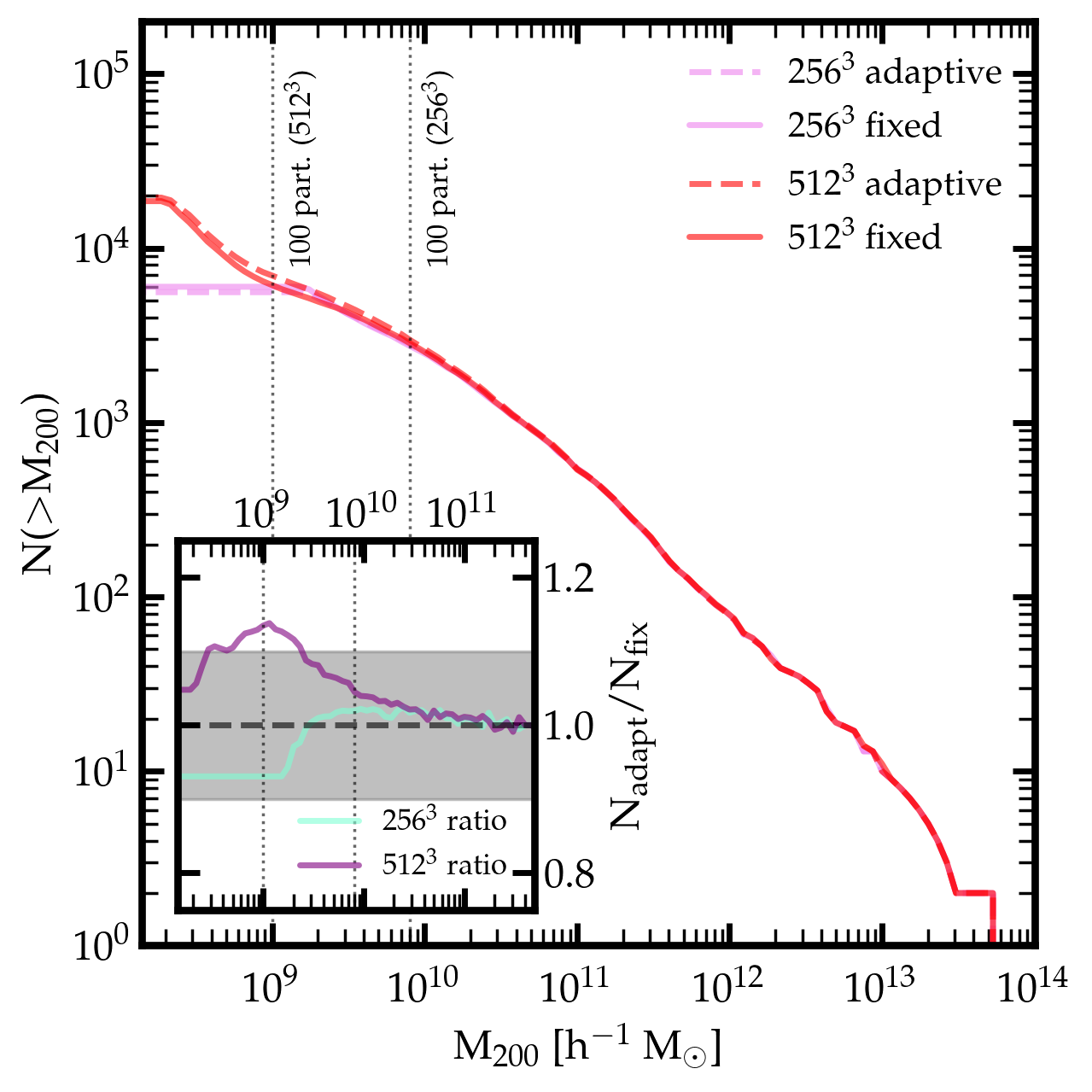}
  \caption{Similar to \Fig{fig:a1_chmf_res_cdm_compare} but for the warm dark matter runs. The low resolution run is represented in red, whereas the higher resolution run is shown in violet.}\label{fig:a1_chmf_res_wdm_compare}
\end{figure}

To investigate the possible resolution dependency of our results, we run a lower resolution counterpart at $N_{\rm part}^3=256^3$ of each simulation and reproduce each analysis done for the higher resolution run. 

In \Fig{fig:a1_chmf_res_cdm_compare} and \Fig{fig:a1_chmf_res_wdm_compare} we show the cumulative halo mass function of the cold and warm dark matter simulations, run with both tidal adaptive and fixed softening, respectively. We observe good convergence above the halo mass limit of the low resolution runs, i.e. $M_{200}\sim 2\times10^{9}$ \hMsun{}.

When comparing the amount of haloes and subhaloes between the cold dark matter tidal adaptive and fixed softening runs, we identify a similar count ratio excess of objects in the adaptive runs with respect the fixed softening counterparts, with peaks around their corresponding 100 particle limit. In the warm dark matter runs, however, the count ratios trends can be different. For the lower $256^3$ resolution runs, the ratio between tidal adaptive and fixed softening is mostly identical for masses M$_{200}>3\times10^9$ \hMsun{}. In the higher $512^3$ resolution runs, the ratio deviates by up to 1.14 above the fixed counts.

\section{Subhalo sample selection}\label{sec:appendix_b}
\begin{figure}
  \includegraphics[width=\columnwidth]{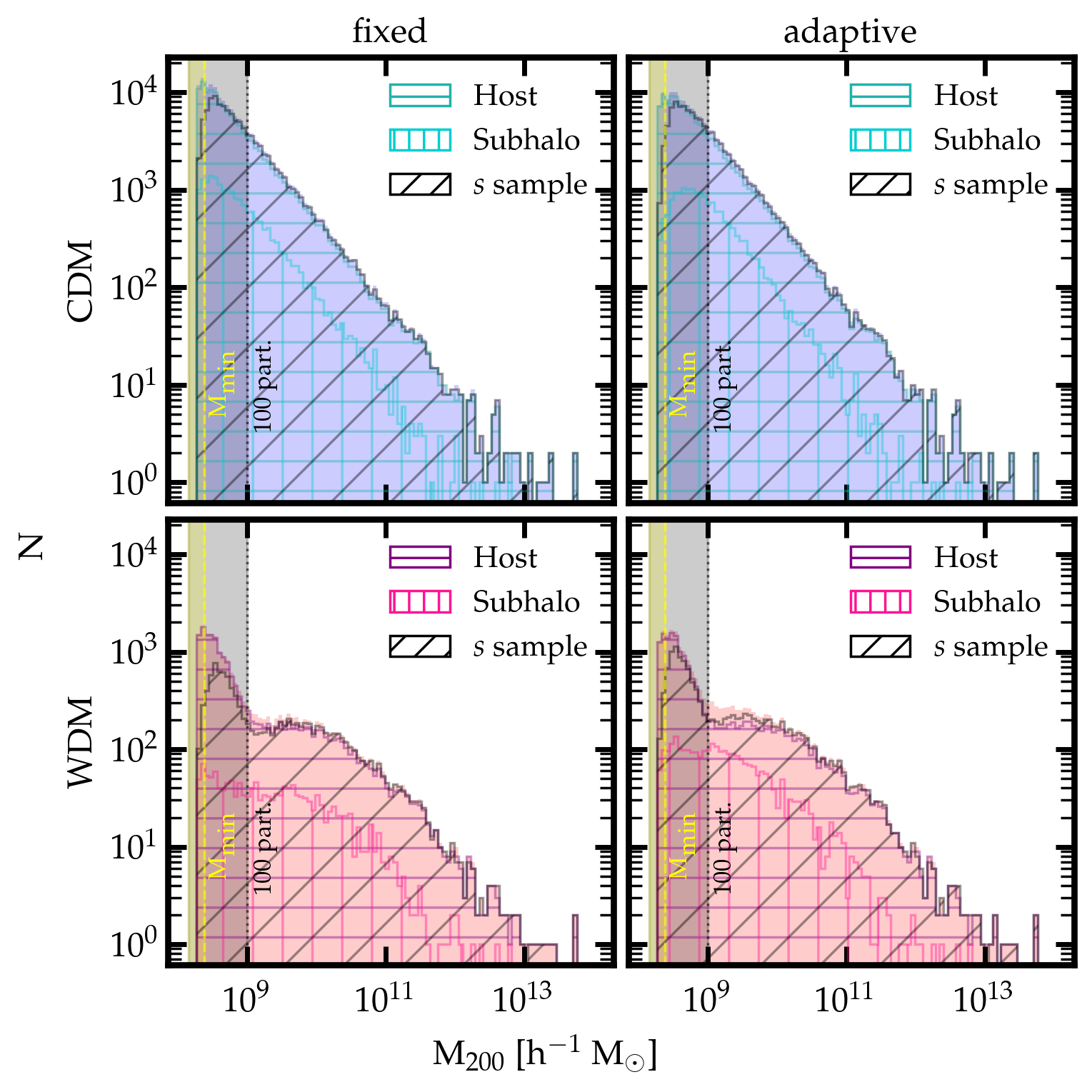}
  \caption{Halo mass distribution of all the self-bound structure identified at $z=0$. The top (bottom) panel corresponds to the cold (warm) dark matter runs. The left (right) column shows the results for the fixed (tidal adaptive) runs. In each panel, the solid colour distribution corresponds to the total sample of objects, including both hosts and subhaloes (i.e. haloes residing within a host halo). The horizontal hatched distribution shows only the host haloes, whereas the lighter vertical hatched distribution represents only the subhalo population. The grey diagonal hatched distribution marks the sub-sample of objects (both hosts and subhaloes) we selected for the sphericity $s$ analysis in \Sec{sec:results-subsec:cosmo}. The yellow vertical dashed line marks the minimum mass M$_{\textrm{min}}$ criterion from \citet{lovell_properties_2014}, whereas the black dotted vertical line marks the mass corresponding to 100 particles.}\label{fig:a2_s_subh_compare}
\end{figure}

In \Fig{fig:a2_s_subh_compare} we show the halo mass distribution of haloes (or "hosts", i.e objects that do not reside within another halo) and subhaloes at $z=0$ for the cold and warm dark matter simulations and for the tidal adaptive and fixed softening runs.  

In each panel we show the total distribution of objects, the individual host and subhaloes distributions, and the sub-sample of objects selected for the mean sphericity analysis. We note that the 100 particle limit (M$_{200} =10^9$ \hMsun{}) captures the population of objects in the upturn of the distribution, whereas the \citet{lovell_properties_2014} M$_{\textrm{min}}$ mass criterion falls below the particle limit.

%%%%%%%%%%%%%%%%%%%%%%%%%%%%%%%%%%%%%%%%%%%%%%%%%%

% Don't change these lines
\bsp	% typesetting comment
\label{lastpage}
\end{document}